\documentclass[aps,prl,twocolumn,superscriptaddress,showpacs]{revtex4-1}
\usepackage{graphicx}
\usepackage{psfrag}
\usepackage{ae}
\usepackage{amsmath,amssymb}
\usepackage{float}

\begin{document}
\date{\today}

\title{Non-equilibrium Statistical Mechanics of Two-dimensional Vortices}

\author{Renato Pakter}
\author{Yan Levin}
\affiliation{
  Instituto de F\'{\i}sica, UFRGS, 
  Caixa Postal 15051, CEP 91501-970, Porto Alegre, RS, Brazil  }


%

\begin{abstract}
It has been observed empirically  that two dimensional vortices tend to cluster forming a giant
vortex. To account for this observation Onsager introduced a concept of negative absolute temperature
in equilibrium statistical mechanics.  In this Letter we will show that in the thermodynamic
limit a system of interacting vortices does not relax to the thermodynamic equilibrium, but becomes trapped
in a non-equilibrium stationary state.  We will show that the vortex distribution in this non-equilibrium stationary state has a characteristic core-halo structure, which can be predicted {\it a priori}.  All the theoretical results are compared with explicit molecular dynamics simulations. 
\end{abstract}

\pacs{05.20.-y, 05.45.-a, 05.70.Ln}

\maketitle

Seventy years ago Onsager presented his celebrated theory of large scale vortex formation in two dimensional turbulence, which for the first time introduced the notion of negative temperature in physics~\cite{On49}. 
Onsager worked in the framework introduced earlier by Helmholtz~\cite{He67} and Kirchhoff~\cite{Ki83} in which the solution to
the incompressible 2d Euler equation is written in terms of a pseudo scalar vorticity 
$\Gamma({\bf r}, t)=[ \nabla \times {\bf u}({\bf r}, t)] \cdot \hat{\bf z}$, 
where ${\bf u}({\bf r}, t)$ is the velocity of fluid at position ${\bf r}$, and $\hat{\bf z}$ is the unit vector normal to the fluid plane.  The incompressibility condition for the Euler equation allows one to introduce a 
stream function $\varphi({\bf r}, t)$ such that 
$\bf u({\bf r}, t)=\nabla \times \varphi({\bf r}, t)\hat{\bf z}$, which satisfies the Poisson equation $\nabla^2 \varphi({\bf r}, t)=-\Gamma({\bf r}, t)$,
the solution to which can be written in terms of an appropriate Green function,
\begin{eqnarray}
\label{e1}
\varphi({\bf r}, t)=\int G({\bf r}, {\bf r'}) \Gamma({\bf r'}, t) d {\bf r'}.
\end{eqnarray}
In an open space the Green function corresponds to the 2d Coulomb-like potential $G({\bf r}, {\bf r'})=-\frac{1}{2 \pi} \ln \vert {\bf r}- {\bf r'} \vert$. Furthermore, it is easy to show that the vorticity field is simply advected by the flow, $d \Gamma({\bf r}, t)/dt=0$.  If we suppose that the vorticity field is composed of various point vortices 
$\Gamma({\bf r})=\sum \Gamma_i \delta ({\bf r} -{\bf r}_i(t))$, their velocity is then 
the same as of the fluid, $\dot{\bf r_i}=\nabla_i \times \sum_{j \ne i} \Gamma_j G({\bf r}_i, {\bf r}_j) \hat{\bf z} $,
and the vortex dynamics has a Hamilton-like structure
\begin{eqnarray}
\label{e3}
\Gamma_i  \dot{x_i} =\frac{\partial \cal H}{\partial y_i} \,\,\,;\,\,\, \Gamma_i  \dot{y_i} = -\frac{\partial \cal H}{\partial x_i}\,.
\end{eqnarray}
The Kirchhoff function is ${\cal H}= \sum_{i<j} \Gamma_i \Gamma_j G(\bf r, \bf r')$, 
where we have removed the singular term, and the 
$x$ and $y$ coordinates of a vortex are the conjugate variables. 
Besides the total energy, the system Eqs.(\ref{e3}), has two other invariants corresponding to the conservation of the total linear and angular momentums of the fluid,
\begin{eqnarray}
\label{e4}
{\bf P} =\sum_i \Gamma_i {\bf r_i } \,\,\,;\,\,\, L=\sum_i \Gamma_i r_i^2\,.
\end{eqnarray}

Onsager's argument for formation of large scale vortex structures is beautiful in its simplicity~\cite{On49}.  Suppose that 
that $N$ vortices are confined in a bounded region of area $A$. Onsager suggested that in the thermodynamic limit,
$N \rightarrow \infty$, Boltzmann-Gibbs statistical mechanics can be applied to the vortex fluid.  The maximum entropy state would then correspond to a completely disordered vortex gas occupying uniformly all of the area $A$.  The energy of this fully disordered state, $E_c$, can be easily calculated using the appropriate Green function.  This means that if $E>E_c$ any inhomogeneous vortex distribution will have lower entropy than $S(E_c)$, so that entropy $S(E)$ will be a decreasing function of energy.  Since the temperature is $1/T=\partial S/\partial E$  
the vortex gas with energy  $E>E_c$  will have negative temperature.  In equilibrium, the probability of a given 
vortex configuration is proportional to the Boltzmann weight --- a negative temperature state~\cite{FrWa15}, therefore, would imply clustering of vortices of the same sign, which would then 
explains spontaneous appearance of large scale vorticity in 2d turbulence.  

Onsager's theory relies on two fundamental assumptions  -- existence of thermodynamic limit and ergodicity of vortex motion.  Because of the long range interaction, the usual thermodynamic limit --- $N \rightarrow \infty$, $A\rightarrow \infty$ with $N/A$ constant --- is not appropriate except for systems with equal number of cyclonic and anti-cyclonic vortices, in which case it was proven rigorously that
the critical energy is infinite, and the temperature is always positive~\cite{FrRu82}.  The interesting case is then a non-neutral system, in particular, the one in which there are only vortices of one sign, and which for simplicity we will assume all to have the same vorticity $\Gamma$.  In this case, the appropriate thermodynamic limit is $N \rightarrow \infty$, $\Gamma \rightarrow 0$, with $\Omega=\Gamma N$ remaining constant~\cite{LuPo77,EySp93}.  In this limit the correlations between vortices vanish and the mean field Poisson-Boltzmann equation becomes exact~\cite{Le02}. Unlike the one component plasmas (OCP) -- which due to repulsion between the particles must be confined by an external potential -- the vortices are ``self-confining" ~\cite{EySp93} because of the conservation  of angular momentum of the fluid, Eq. (\ref{e4}), which acts as an effective external potential. It is convenient to define the effective vortex charge $q=\Gamma/\sqrt{2 \pi}$ so that the interaction potential between the vortices becomes identical to that of charges in a two dimensional one component plasma. In equilibrium, the ``electrostatic" potential $\psi=\varphi \sqrt{2 \pi}$ will then satisfy the 2d Poisson-Boltzmann (PB) equation,
\begin{eqnarray}
\label{e5}
\nabla^2 \psi=-2 \pi q e^{-\beta \psi -\beta \alpha r^2 -\beta \mu } \,,
\end{eqnarray}
where $\beta=1/k_B T$, $\alpha$, and $\mu$ are respectively the Lagrange multiplier for the conservation of energy, angular momentum, and the total vorticity~\cite{LuPo77,ChSo96,BoVe12}.  Starting with an initial vortex distribution, the equilibrium distribution can be calculated by numerically solving the non-linear Poisson-Boltzmann (PB) equation with the boundary conditions $\psi(0)=0$
and $\psi'(0)=0$ and requiring that asymptotically the potential goes as $\psi(r) \sim -q N \ln(r)$.

\begin{figure}
\begin{center}
\includegraphics[scale=.5]{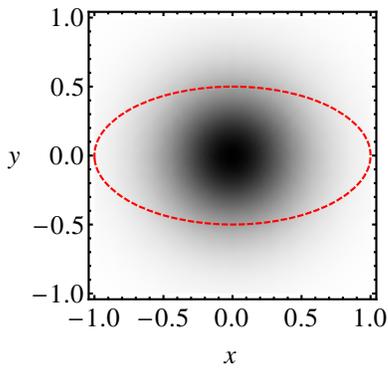}
\caption{The equilibrium vortex density distribution starting from an initial state in which vortices are uniformly distributed inside an ellipse with
$a=1.0$ and $b=0.5$, shown by dashed curve, 
calculated by solving numerically the non-linear PB equation (\ref{e5}). The equilibrium distribution has negative temperature corresponding to $\beta=-1.19$}
\label{veq}
\end{center}
\end{figure}

\begin{figure}
\begin{center}
\includegraphics[scale=.35]{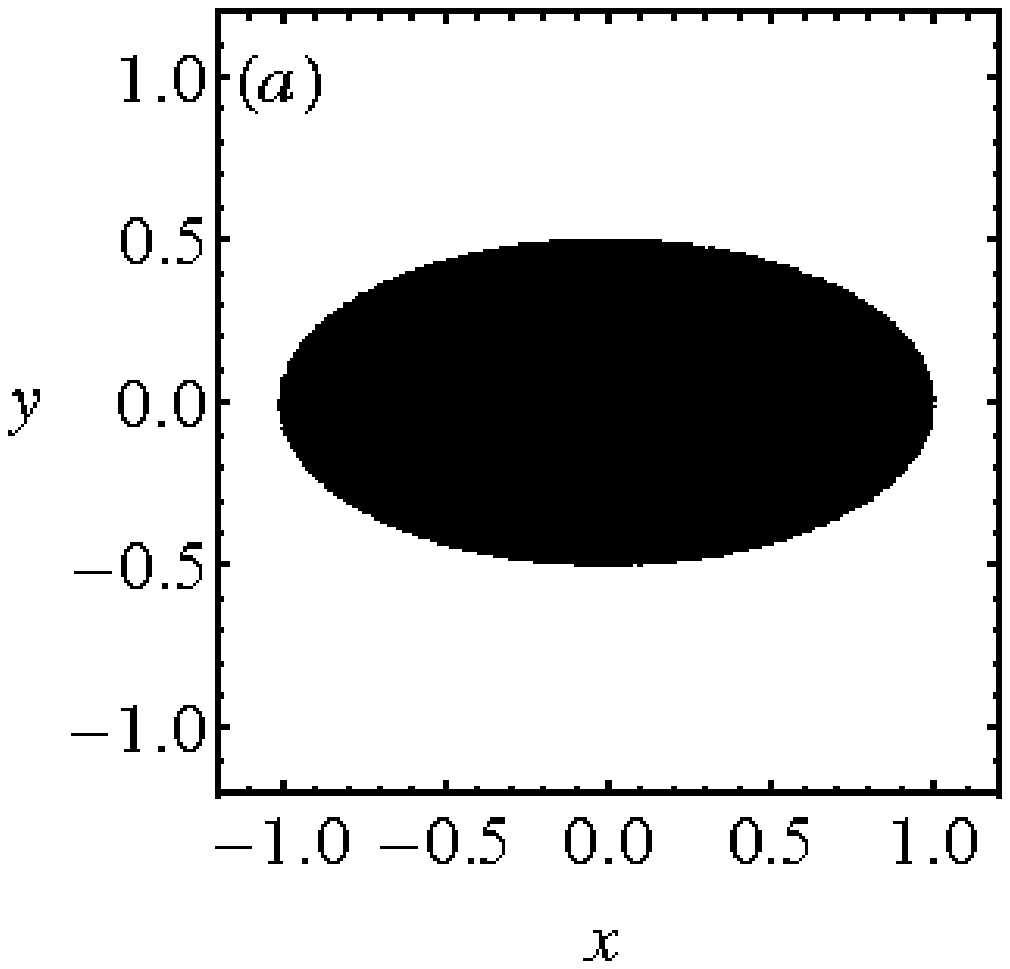}
\includegraphics[scale=.35]{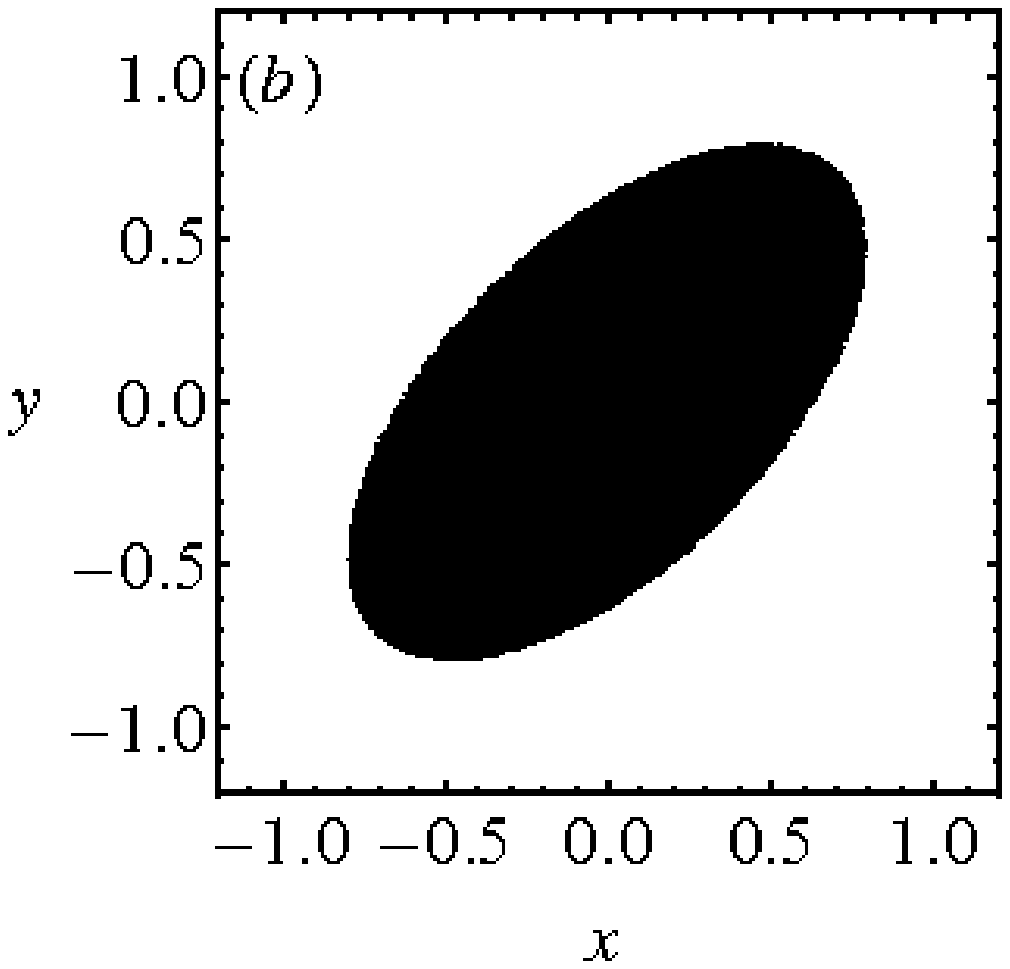}
\includegraphics[scale=.35]{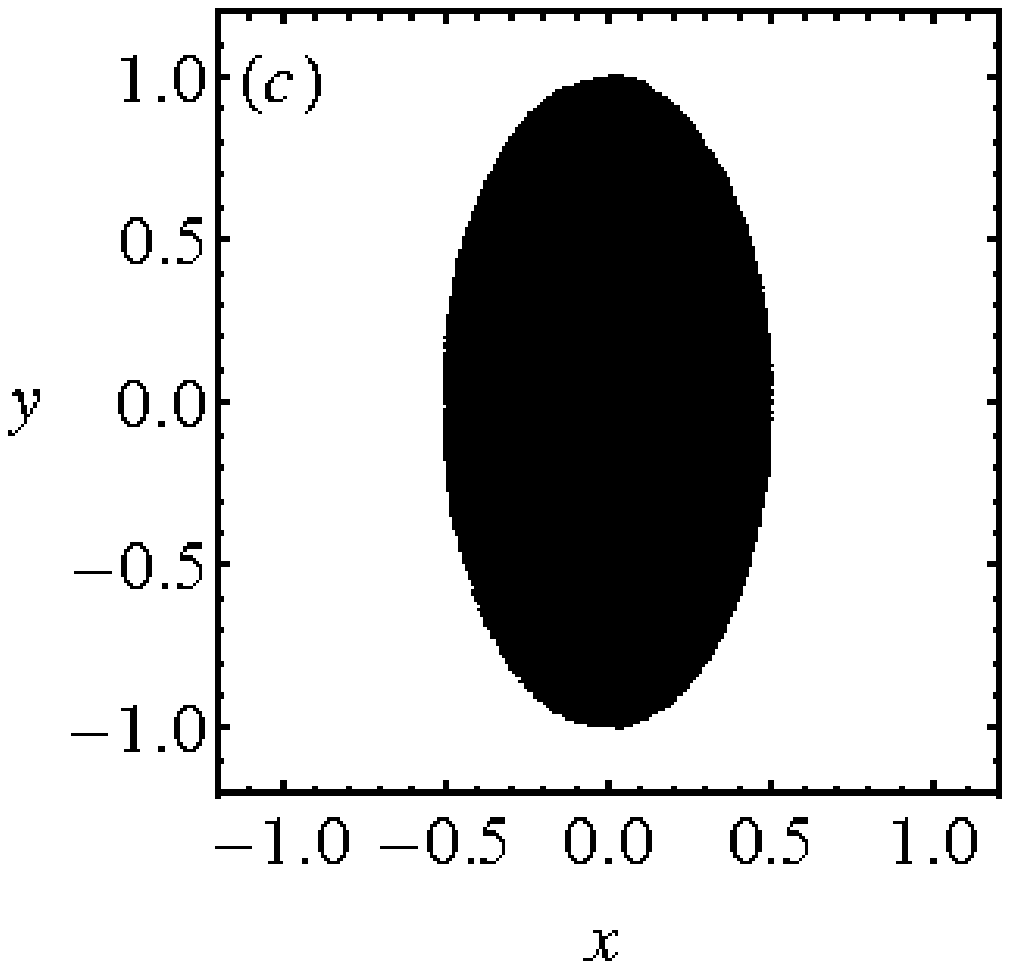}\hspace{.2cm}
\includegraphics[scale=.34]{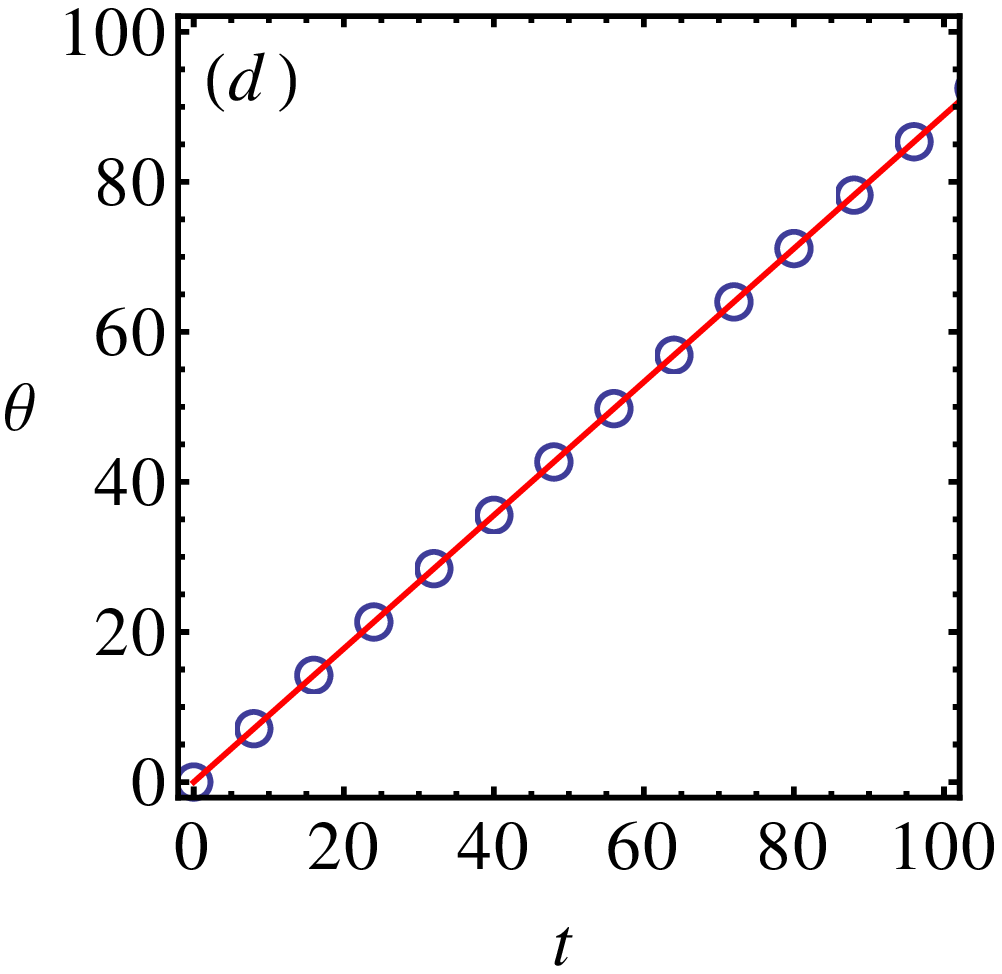}
\caption{Snapshots of molecular dynamics simulation 
for elliptical vortex distribution with   
$a=1.0$ and $b=0.5$ and $N=10^6$ at: (a) $t=0$, (b) $t=T/8$ , and (c) $t=T/4$, where $T=2\pi/\omega$ and $\omega$ is given by Eq.(\ref{omega}). In (d) we show the time evolution of the angle between semi-major axis of ellipse and the  $x$-axis, $\theta$.  The circles correspond to the results obtained from MDS and the line is the theoretical prediction, $\theta=\omega t$.}
\label{vlequil}
\end{center}
\end{figure}

Consider an initially uniform elliptical distribution of vortices
\begin{equation}
f_{ell} (x,y)=\eta \, \Theta\left[1-{x^2\over a^2}-{y^2\over b^2}\right ] ,
\label{fell}
\end{equation}
where $\eta=N/\pi a b$ and $\Theta$ is the Heaviside step function. Solving numerically the PB equation we find that Onsager's theory predicts that this initial distribution will relax to a spherically symmetric equilibrium state with negative temperature depicted in Fig. \ref{veq}. 
To check the validity of Onsager's theory we performed N-body molecular dynamics simulations (MDS) using a particle-in-cell (PIC) algorithm \cite{ChZa73,DrSc08,DrSc09} with an adaptive time-step integrator that uses an embedded fifth and sixth order
Runge-Kutta estimates to calculate vortex trajectories and the relative errors to adjust the step size \cite{NumRecipesC}.  This  significantly speeds up the MDS time. Alternatively one could also use a symplectic integrator \cite{DuFr07,DuFr10}.  A PIC algorithm is particularly useful for vortex simulations since it eliminates the collisional finite size effects which
are present in direct pairwise-interactions MDS, but which must vanish in the 
thermodynamic limit \cite{LePa14}.
Starting with
the initial elliptical vortex distribution, Eq. (\ref{fell}), with 
$a=1.0$ and $b=0.5$,  we simulated the dynamics of $N=10^6$ vortices.
The snapshots of various temporal configurations are presented in panels (a) to (c) of Fig. \ref{vlequil}.  The figure shows that instead of relaxing to equilibrium, the initial particle distribution undergoes a rigid rotation with a constant angular velocity, maintaining its  elliptical shape and uniform density, see Fig. \ref{vlequil}. 


To understand the discrepancy between the simulations and Onsager's theory we must turn to kinetic theory.  In the thermodynamic limit  --- $N \rightarrow \infty$, $q \rightarrow 0 $ and $q N=1$  ---  the evolution of the vortex distribution function is governed { \it exactly}~\cite{BrHe77} by the Vlasov
equation,
\begin{equation}
 \frac{\partial f}{\partial t}+\frac{\partial \psi}{\partial y} \frac{\partial f}{\partial x} -
\frac{\partial \psi}{\partial x} \frac{\partial f}{\partial y}=0 \,.
\end{equation}
Vlasov equation is identical to the condition that vortices are advected by the flow, $d \Gamma({\bf r}, t)/dt=0$, so that the vortex gas evolves as an incompressible fluid.  

The electrostatic potential for an elliptical distribution Eq. (\ref{fell}) can be calculated explicitly~\cite{PaLe07,Ke53}
\begin{equation}
 \psi_{ell}(x,y)=
\begin{cases}
\psi_{in}(x,y),&{\rm for}\ (x/a)^2+(y/b)^2\leq 1,\\
\psi_{out}(x,y),&{\rm for}\ (x/a)^2+(y/b)^2> 1,
\end{cases}
\end{equation}
where
\begin{equation}
 \psi_{in}(x,y)=\log \left(\frac{2}{c}\right)-\cosh
   ^{-1}\left(\frac{a}{c}\right)-\frac{x^2}{a
   (a+b)}-\frac{y^2}{b (a+b)}+\frac{1}{2},
   \label{in}
\end{equation}
\begin{eqnarray}
&&\psi_{out}(x,y)=\log \left(\frac{2}{c}\right)+ \nonumber \\
&& \Re e \left [\frac{z^2}{c^2}\left(\sqrt{1-\frac{c^2}{z^2}}-1\right)-\cosh
   ^{-1}\left(\frac{z}{c}\right)\right ]+\frac{1}{2},
   \label{out}
\end{eqnarray}
and $c=\sqrt{a^2-b^2}$, $z=x+iy$, $i=\sqrt{-1}$, and $\Re e$ stands for the real part of the expression. The potential and its derivatives
are continuous along the boundary of the ellipse, and asymptotically for large distances, $\psi_{out}\to -\ln(r)$.

We now observe that a given distribution corresponds to Vlasov equilibrium if it depends on the phase space variables only through the conserved quantities. Since the potential (stream function) plays the role of a Hamiltonian for one particle dynamics, if the distribution function would  depend on $x$ and $y$ only through the equilibrium potential $\psi(x,y)$, then $f(\psi(x,y))$  would be Vlasov stationary.   A direct inspection of the potential inside an ellipse, Eq.~(\ref{in}), however, shows that the equipotentials are ellipses of semi-radii proportional to $\sqrt{a(a+b)}$ and $\sqrt{b(a+b)}$, which are different from those of the initial vorticity distribution, $a$ and $b$, respectively.  Hence, the boundary of the initial distribution is not an equipotential and the initial elliptical distribution will evolve in time.   

Let us now consider a rotating ellipse with $f(x,y,t)=f_{ell} (\tilde x,\tilde y)$, where
\begin{eqnarray}
\nonumber
 \tilde x &=& x\cos(\omega  t)+y\sin (\omega  t), \\
 \tilde y &=& -x\sin(\omega  t)+y\cos (\omega  t),
 \label{transf}
\end{eqnarray}
and $\omega $ is some angular velocity. The dynamics in the rotating reference frame can be studied using a canonical transformation with a generating function
\begin{equation}
 {\cal F}(x,\tilde y)={x\tilde y\over \cos(\omega t)}+{x^2+\tilde y ^2\over 2} \tan (\omega t),
\end{equation}
such that $\tilde x=\partial {\cal F}/\partial \tilde y$ and $y=\partial {\cal F}/\partial x$ correspond to Eqs. (\ref{transf}). Since the generating
function depends explicitly on time, the effective interaction potential in the rotating reference frame is
$\tilde \psi_{ell}=\psi_{ell}+\partial {\cal F}/\partial t$, which reduces to
\begin{equation}
\tilde \psi_{ell}(\tilde x,\tilde y)=\psi_{ell}(\tilde x,\tilde y)+{\omega (\tilde x^2+\tilde y^2)\over 2}.
\label{psit}
\end{equation}
The question now is can we find a frequency $\omega$ such that the boundary of the distribution is an equipotential of $\tilde \psi_{ell}$? Substituting Eq.~(\ref{in}) into Eq.~(\ref{psit}) and evaluating the potential
at the  boundary of the ellipse, $(\tilde x/a)^2+(\tilde y/b)^2= 1$, we see that the potential will be constant (independent of $\tilde x$ and $\tilde y$ along the boundary) if 
\begin{equation}
\omega=\frac{2}{(a+b)^2}\,.
\label{omega}
\end{equation}
Hence, a uniformly distributed ellipse can be written in terms of the single particle conserved 
quantity $\tilde \psi_{ell}(\tilde x,\tilde y)$ as $f_{ell} (\tilde x,\tilde y)=\eta \,\Theta\left[\tilde \psi_{ell}^<(\tilde x,\tilde y)-\epsilon_{ell}\right],$
where $\epsilon_{ell}=-ab/(a+b)^2$ is the constant effective potential along the ellipse boundary.  The distribution $f_{ell}$ is, therefore, a Vlasov equilibrium in the frame that rotates with angular velocity given by Eq.~(\ref{omega}).  
The effective potential in Eq.~(\ref{psit})
is a nonmonotic function, presenting a local maximum at the origin, extremum  curve connecting the inflection points, and diverging as $\tilde r \rightarrow \infty$.  Therefore, we use the superscript ``$<$'' to indicate that we are considering the inner (between the origin and the extremum curve) branch of $\tilde \psi_{ell}(\tilde x,\tilde y)$. The rotation velocity $\omega$ is precisely the
one that was found in our molecular dynamics simulations, Fig. \ref{vlequil}d.  This explains why the initial vortex distribution does not
relax to Onsager predicted equilibrium, but instead rotates as a rigid object.  

\begin{figure}
\begin{center}
\includegraphics[scale=.4]{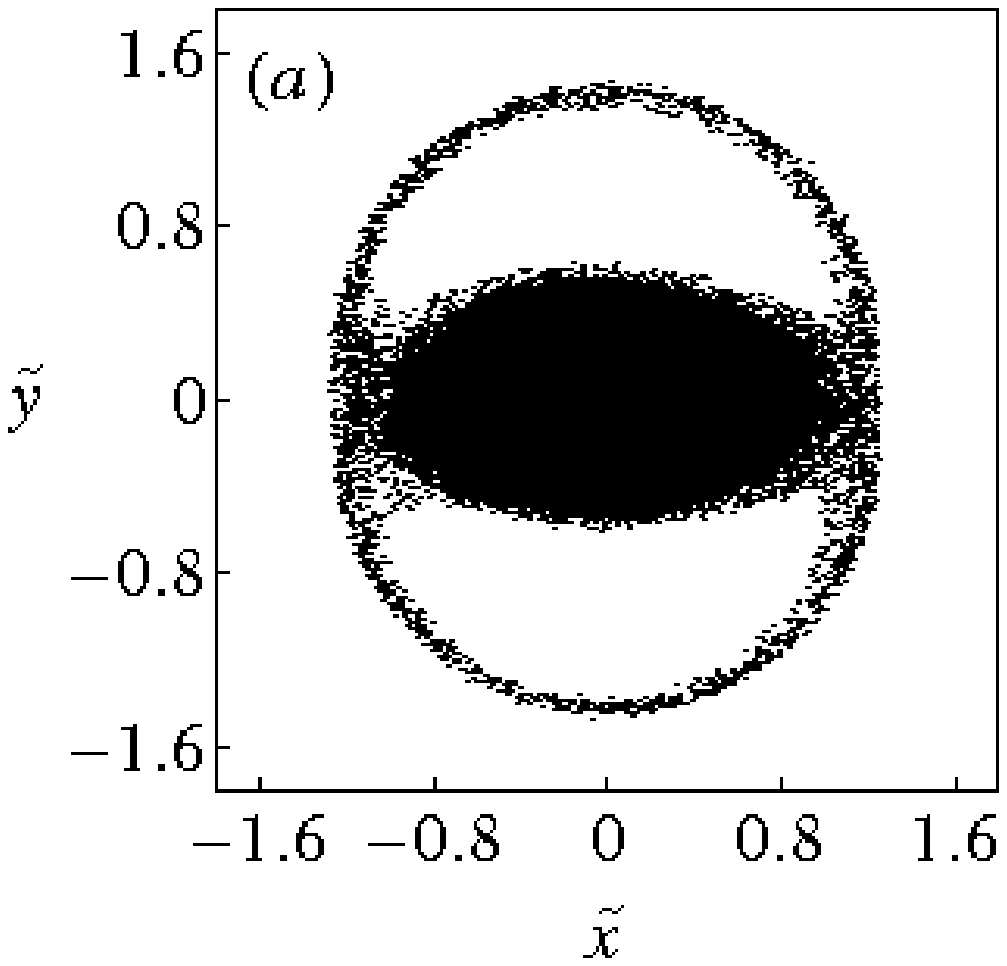}
\includegraphics[scale=.4]{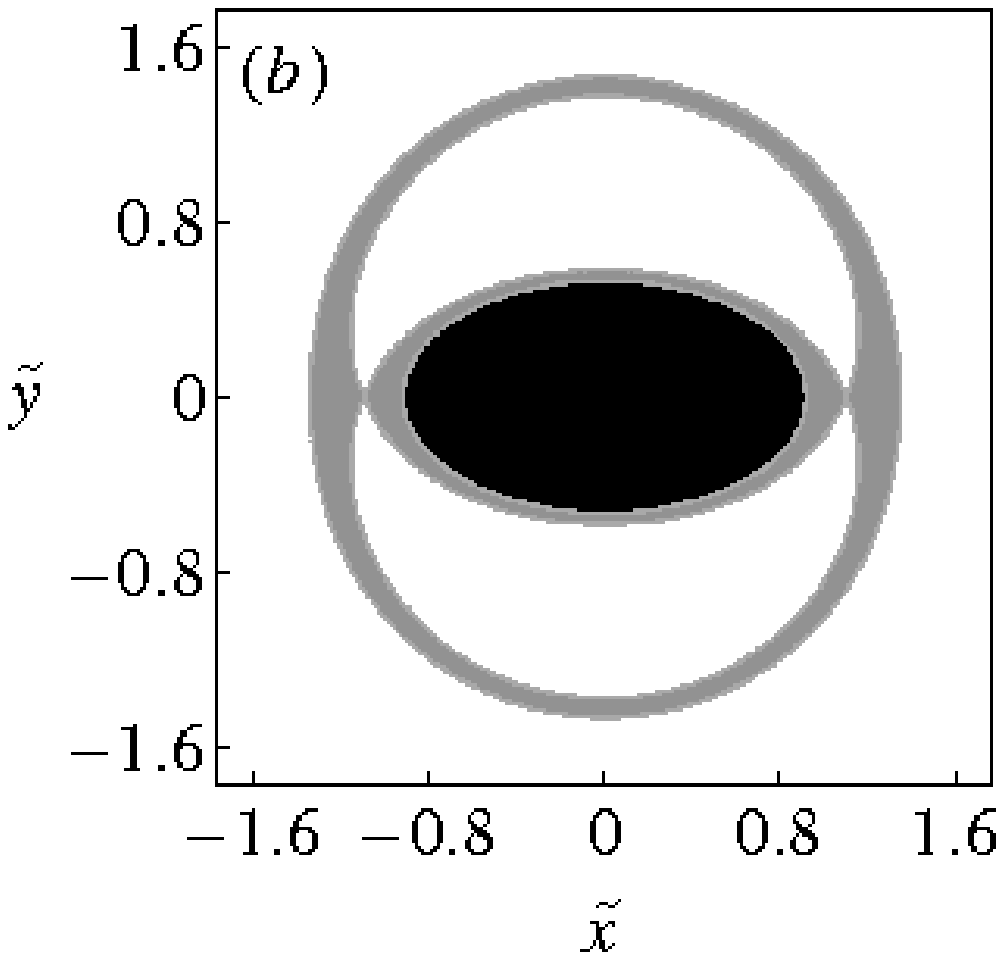}
\caption{(a) Snapshot of the phase space obtained using molecular dynamics simulation. (b) The theoretical prediction obtained using Eq.~(\ref{fch}), with no adjustable parameters.  The black region corresponds to the high density core, whereas the 
gray region corresponds to the low density halo. The core-halo structure rotates in the lab frame. The core has population inversion  -- the high energy states are occupied up to the maximum density $\eta$ permitted by the Vlasov dynamics.}
\label{corehalo}
\end{center}
\end{figure}

The next question to address is if the Vlasov 
equilibrium state is stable. That is, if the  initial elliptical distribution is 
perturbed, will it then relax to Onsager equilibrium?  Based on non-equilibrium 
statistical mechanics of systems with long range interactions~\cite{CaDa09,LePa14,CaDa14} we do not expect this to be the case and instead expect that the system will relax, in a coarse grained sense~\cite{PaLe17}, to a core-halo structure observed in magnetically confined plasmas~\cite{LePa08}, gravitational systems~\cite{TeLe10,TeLe11, JoWo11}, and spin models~\cite{PaLe11, Ro16}. 
This is precisely what is found in simulations,
see Fig. \ref{corehalo}.  

\begin{figure}
\begin{center}
\includegraphics[scale=.5]{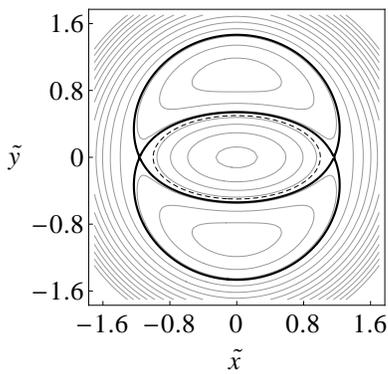}
\caption{ Level curves -- in the rotating reference frame -- of the effective potential,  Eq.~(\ref{psit}), generated by a uniform ellipse rotating with angular velocity $\omega$. The dashed curve shows the ellipse boundary with $a=1.0$ and $b=0.5$. The thick curve corresponds to the separatrix that
contains two hyperbolic fixed points located at $\tilde x=\pm\tilde x_{fix}$ and $\tilde y=0$.}
\label{equi}
\end{center}
\end{figure}

To understand the core-halo distribution observed in simulations we begin by considering the dynamics of  
a test vortex interacting with a rotating ellipse of a uniform vortex density, $f(x,y,t)=f_{ell} (\tilde x,\tilde y)$.  The electrostatic potential produced by such ellipse  is given
by Eq. (\ref{psit}). The equipotentials of $\tilde \psi_{ell}$ correspond to the trajectories of test vortices in the rotating reference frame. 
An example of such equipotentials are shown in Fig. \ref{equi} for $a=1.0$ and $b=0.5$. We notice
a separatrix of a resonant structure (thick solid curve) which can drive test vortices that are just outside the elliptical distribution to large radii. The separatrix
presents two hyperbolic fixed points along the $\tilde x$ axis. Since ${\partial\tilde \psi_{ell}/ \partial \tilde y}=0$ is automatically
satisfied along the $\tilde x$ axis, the position of the fixed point is determined by imposing
${\partial\tilde \psi_{ell}/ \partial \tilde x}|_{(\tilde x,\tilde y)=(\tilde x_{fix},0)}=0$,
which leads to $\tilde x_{fix}=\pm \left [{(a+b)^3\over a+3b}\right]^{1/2}$.
Hence, the separatrix in the phase space corresponds to all points that satisfy $\tilde \psi_{ell}(\tilde x,\tilde y)=\tilde \psi_{ell}(\tilde x_{fix},0)$. The maximum radius achieved along the
separatrix -- which occurs at $\tilde x=0$ -- can be computed by solving $\tilde \psi_{ell}(0,\tilde y_{max})=\tilde \psi_{ell}(\tilde x_{fix},0)$ for $\tilde y_{max}$. For the case shown in Fig. \ref{equi}, $\tilde y_{max}\approx 1.5$.
Note that since the separatrix is outside the elliptical distribution, we need to take into account the 
corresponding potential given by Eq. (\ref{out}) in the derivations.

The dynamical mechanism behind the core-halo halo formation is now clear.  The parametric resonances capture some vortices and expel them into the {\it low energy} phase space region, far from the main core.  To conserve the total energy of the system, the other vortices must then compensate and move into the high energy core region creating a population inversion.  However, because of the  incompressibility of the Vlasov dynamics, the core density can not exceed  $\eta$ determined by the initial distribution function Eq. (\ref{fell}). The resonant mechanism of vortex evaporation will then lead to formation of a high energy core region in which all the energy states up to the ``Fermi energy" $\epsilon_F$ are fully occupied with maximum allowed density $\eta$.  The stationary distribution will be established when the rates of evaporation and condensation become identical. 
We now propose an ansatz solution for the Vlasov stable stationary distribution function in the rotating reference frame which has a core-halo form,
\begin{eqnarray}
&& f_{ch}(\tilde x,\tilde y)=\eta\, \Theta\left[\tilde \psi_{ell}^<(\tilde x,\tilde y)-\epsilon_F\right]+\nonumber \\
&&\chi \, 
 \Theta\left[\epsilon_h-\tilde \psi_{ell}^>(\tilde x,\tilde y)\right] \Theta\left[\tilde \psi_{ell}(\tilde x,\tilde y)-\epsilon_{sep}\right] \times \nonumber \\
&& \Theta\left[\epsilon_F-\tilde \psi_{ell}^<(\tilde x,\tilde y)\right],
 \label{fch}
\end{eqnarray}
where $\chi$ is the halo density, 
$\epsilon_F=\tilde \psi_{ell}^<(a_s,0)=\tilde \psi_{ell}^<(0,b_s)$, 
$\epsilon_{sep}=\tilde \psi_{ell}(\tilde x_{fix},0)$, $\epsilon_h=\tilde \psi_{ell}(0,\tilde y_{max})$, 
$\tilde \psi_{ell}(\tilde x,\tilde y)$ is approximated as the effective potential created by the core of the distribution which corresponds to 
an ellipse of semi-radii $a_s$ and $b_s$, $y_{max}$ is the halo size computed from the separatrix of the initial ellipse, and the $>$ ($<$)
superscript indicate that we consider solely the outer (inner) branch of the effective potential. 
The core-halo distribution of Eq.~(\ref{fch}) has 3 unknown parameters --- the semi-radii of the final stationary elliptical core, $a_s$, $b_s$, and the halo density $\chi$. These parameters can be determined by imposing the conservation of the total vorticity, of the total energy, and of the angular momentum $L$,
\begin{eqnarray}
&&\int d^2 {\bf r} f_{ch}({\bf r}) = N \nonumber \\
&&\frac{q^2}{2} \int d^2 {\bf r} \, d^2 {\bf r}' f_{ch}({\bf r}) f_{ch}({\bf r}') \ln\vert {\bf r} -{\bf r'} \vert=  \nonumber \\
&&\frac{1}{8} \left[1-4 \ln \left(\frac{a+b}{2}\right)\right] \nonumber \\
&&\int d^2 {\bf r} \, r^2 \, f_{ch}({\bf r}) =\frac{N}{4} \left(a^2+b^2\right) \,,
\end{eqnarray}
respectively.
Note that $\bf P$ is automatically conserved because of the symmetry of the distributions with respect to the origin that guarantees that $\langle x \rangle=0$ and $\langle y \rangle=0$.
In Fig. \ref{corehalo} we compare the stationary state obtained using molecular dynamics simulation with the theoretical solution given by Eq.~(\ref{fch}). An excellent agreement is found between the two, without any adjustable parameters.

We have presented a theory which accounts for the relaxation of an initial vortex distribution to the final stationary  -- in the rotating reference frame --- state.  Contrary to Onsager's theory, the initial distribution does not
relax to thermodynamic equilibrium with symmetric vortex distribution and negative temperature.  Instead we find that the system evolves to a complicated non-rotationally symmetric core-halo structure
which rotates at a constant frequency in the lab frame. As suggested by Onsager, we find that the distribution corresponds to the population inverted state in which the high energy states are occupied up to the maximum density permitted by the incompressibility condition of the Vlasov dynamics.

There is a profound difference between the vortex dynamics and that of a one component plasma confined by magnetic field~\cite{Gl94}.  In both cases the system is observed to relax to a core-halo distribution. In the case of plasmas, however, resonances lead to particle evaporation and the condensation of remaining charges into the {\it lowest energy} states through the process of Landau damping~\cite{La46}, leading to a stationary core-halo distribution function in the {\it lab frame}~\cite{LePa08}.  In the case of vortex dynamics, the situation is reversed, and resonances result in a population inversion such that high energy core region is occupied up to the maximum allowed phase space density. Furthermore, the stationarity is achieved only in the rotating reference frame.  The population inversion of the core region may be associated with the negative temperature proposed by Onsager. However, since in the thermodynamic limit the vortex gas always remains out of equilibrium, the temperature is not a well defined concept in this context. Finally, it should be interesting to extend our result to the quantum 
regime in which vortex condensates have also been observed~\cite{SuDa14, GrDa18}.

Y.L. would like to thank Paul Wiegmann for bringing this problem to his attention.
This work was partially supported by the CNPq, National Institute of Science and Technology Complex Fluids INCT-FCx, and by the US-AFOSR under the grant 
FA9550-16-1-0280.

\end{document}